# Efficient energy, cost reduction, and QoS based routing protocol for wireless sensor networks

**Ghassan Samara, Mohammad Aljaidi**
Computer Science Department Zarqa University, Jordan

**ABSTRACT**

Recent developments and widespread in wireless sensor network have led to many routing protocols, many of these protocols consider the efficiency of energy as the ultimate factor to maximize the WSN lifetime. The quality of Service (QoS) requirements for different applications of wireless sensor networks has posed additional challenges. Imaging and data transmission needs both QoS aware routing and energy to ensure the efficient use of sensors. In this paper, we propose an Efficient, Energy-Aware, Least Cost, (ECQSR) quality of service routing protocol for sensor networks which can run efficiently with best-effort traffic processing. The protocol aims to maximize the lifetime of the network out of balancing energy consumption across multiple nodes, by using the concept of service differentiation, finding lower cost by finding the shortest path using nearest neighbor algorithm (NN), also put certain constraints on the delay of the path for real-time data from where link cost that captures energy nodes reserve, energy of the transmission, error rate and other parameters. The results show that the proposed protocol improves the network lifetime and low power consumption.

*Corresponding Author:*

Ghassan Samara,
Computer Science Department Zarqa University, Jordan.
Email: gsamara@zu.edu.jo

## 1. INTRODUCTION

Wireless Sensor Networks (WSNs) are rapidly emanating as an important and influential factor in mobile computing, wireless systems, and vehicular- ad hoc networks [1], Also, WSNs play a major role in facilitating the work of applications in a wide range of areas, such as earthquake monitoring, data traffic in transportation, tracking goals in battlefields, habitat monitoring, fire system detection [2, 3, 4].

Wireless Sensor Networks (WSNs) may have millions of nodes, which are deployed over the wide sensing area in different parameters. These Nodes broadcast their link quality, which is depending on radio frequency environment [5]. The main purposes of WSNs are to monitor, analyze, combine and respond to the data which are gathered by hundreds or thousands of sensors which are distributed in some sensing field [6].

The power of WSNs lies in the capability of deploying significant numbers of tiny nodes that have been configured by themselves, each of these sensing nodes contains a microcontroller, external memory, power source, and transceiver. Figure 1 shows the internal components of a sensor node [1].

In this paper, (ECQSR) protocol proposes; Efficient Energy, Cost Reduction, and QoS based routing protocol for WSNs, to achieve load balancing by splitting the data traffic among set of nodes which are distributed in the field, in addition to efficiently balance the consumption of energy through multiple nodes [7, 8].

Moreover, ECQSR works on increasing the reliability of the data delivery using forward error correction (FEC) technology, also, increasing the throughput via presenting data redundancy, Data-redundancy provides type of flexibility on the path failure, in addition to recover lost data, also it has the possibility of reconstructing the original messages, ECQSR protocol finds the shortest path which has the least cost by using the nearest neighbor algorithm (NN). This path meets the end to end delay requirements [7], Figure 2 shows Pseudocode of the nearest neighbor algorithm (NN).

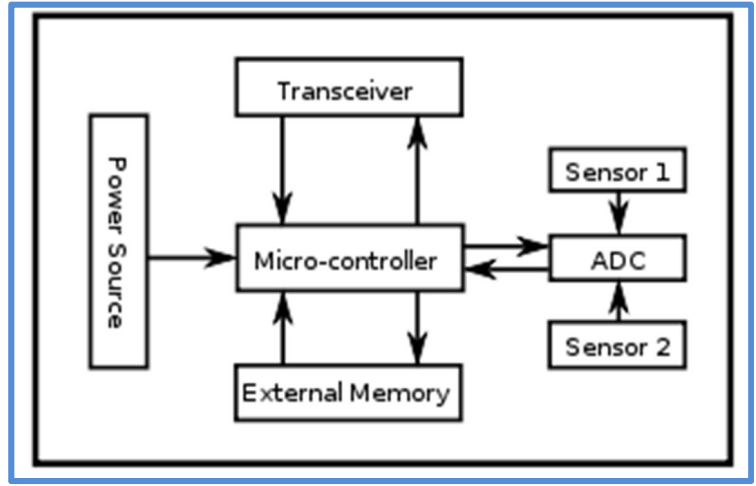

Figure 1. Internal components of a sensor node

```
Input   :  D, a chunk of the original distance matrix;
           n_chunksize, dimension of the chunk;
           split, index of split;
           chunk, index of chunk;
           Maxk, an array to hold the farthest neighbors for each row index in the chunks;
Output:    none, an (intermediate) kNN graph stored in Gk';

1  row' ← blockIdx.x × blockDim.x + threadIdx.x;
2  if row' < n_chunksize then
3      row ← split × n_chunkSize + row' // absolute row in the distance matrix;
4      for column' ← 1 to n_chunksize do
5          column ← chunk × n_chunkSize + column' //absolute column in the distance matrix;
6          if row = column or row > n_row or column > n_col then
7              continue /* exclude diagonal and pad regions */;
8          if D[row' × n_chunksize + column'] < Gk'[Maxk[row']].weight then
9              Gk'[Maxk[row']].source ← row;
10             Gk'[Maxk[row']].target ← column;
11             Gk'[Maxk[row']].weight ← D[row' × n_chunksize + column'];
12         Search the new maximum element in row'(D) and store the index in Maxk[row'];
```

Figure 2. KNN kernel algorithm pseudocode

Figure 3 shows the flow chart of the shortest path which is determined by the NN algorithm.

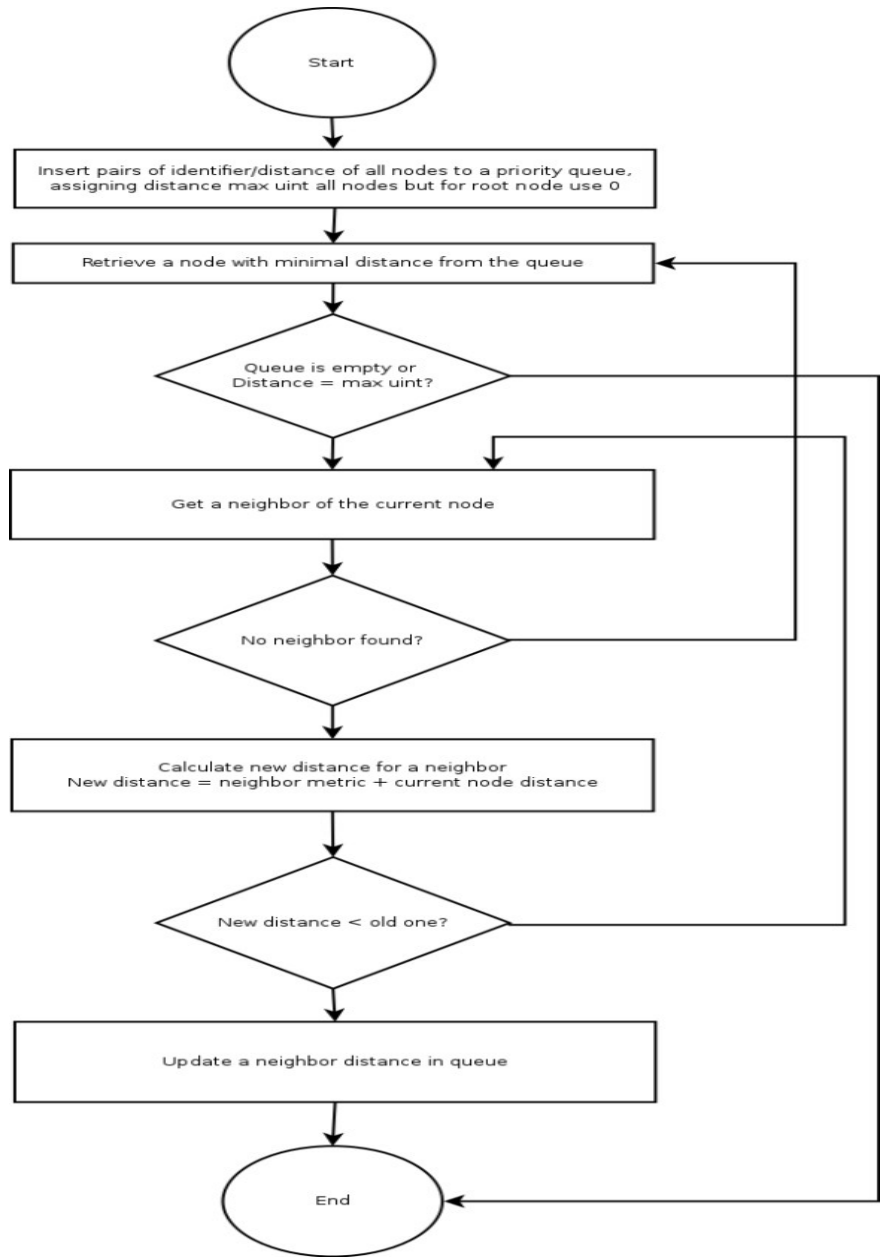

Figure 3. Shortest path flow chart

## 2. RESEARCH PROBLEM
Providing energy efficient, least cost, and QoS based routing protocol between sensor nodes in a particular sensing field is a challenging task, because of power consumption [9, 10, 11], data loss, topology control, coverage, mobility, routing and other parameters [1]. All of these parameters are considered in the proposed protocol. This protocol will work on finding an optimal solution for all of these challenges so that the proposed ECQSR protocol based on the idea of extending the routing approach and determining the path with least cost and energy efficient that satisfy certain end to end delay during transmission [7, 12, 13].

The link cost function is a function that captures transmission energy, reserve energy, error rate, and other parameters. Our protocol aims to get the shortest path with least cost by using the nearest neighbour algorithm (NN), which should meet the end to end delay requirements.

## 3. LITERATURE REVIEW

QoS routing in wireless sensor networks (WSNs) is a challenging problem because of the poor resources of the sensor nodes. Therefore, this issue has received increasing interest from the researchers, where they have made many proposals; here are some of these proposals which relate to our protocol (ECQSR):

Sequential Assignment Routing (SAR) protocol, which is one of the oldest protocols that concern in routing protocols that provide some QoS. SAR is a multiple path routing protocol that helps in routing decisions based on three aspects: QoS on each path, energy resources, and packet priority level. However, SAR protocol suffers from the overhead of dealing with routing tables and QoS metrics at each sensor node [1, 7].

K. Akkaya and M. Younis in proposed a cluster-based QoS aware routing protocol that employs a queuing model to deal with both non-real time and real-time traffic. This protocol suffers from the transmission delay which is not considered in the estimation of the end-to-end delay [1]. SPEED is considered as another protocol of QoS based routing protocol that supports a soft real-time end to end pledges; But the main problem of this protocol, that it does not consider the energy metric in its routing protocol [1, 7].

Message initiate Constrained Based Routing (MCBR) mechanism. MCBR is composed of route constraints, QoS requirement of messages, and specifications of constraint-based destinations, but it suffers from a large number of control packets that yields more overhead [1]. Felemban *et al.* proposed multiple paths and multiple speed Routing Protocol (MMSPEED) for probable QoS guarantee in WSNs, QoS is provided in two domains: reliability domain, and timelines domain, this protocol suffers from dealing with data redundancy [1].

X. Huang, Y. Faun has proposed a multi-constrained QoS multiple path routing (MCMP) protocol, the objective of this protocol is to utilize the multi-path to maximize the performance of the network with a minimum cost of energy, but it does not get over the problem of power consumption completely [1]. Efficient, Least Cost, Energy-Aware (ELCEA) QoS Protocol, which aims to extend the routing and finding an efficient path using Dijkstra's algorithm and least cost with consideration of end-to-end delay during the current connection, but it failed in giving more importance on Quality of Service requirements [1, 7].

## 4. RESEARCH OBJECTIVES

There is a set of objectives, which can be achieved through the use of our protocol (ECQSR), all these objectives must be compatible with each other. Figure 4 shows the relations between these objectives [8, 14].

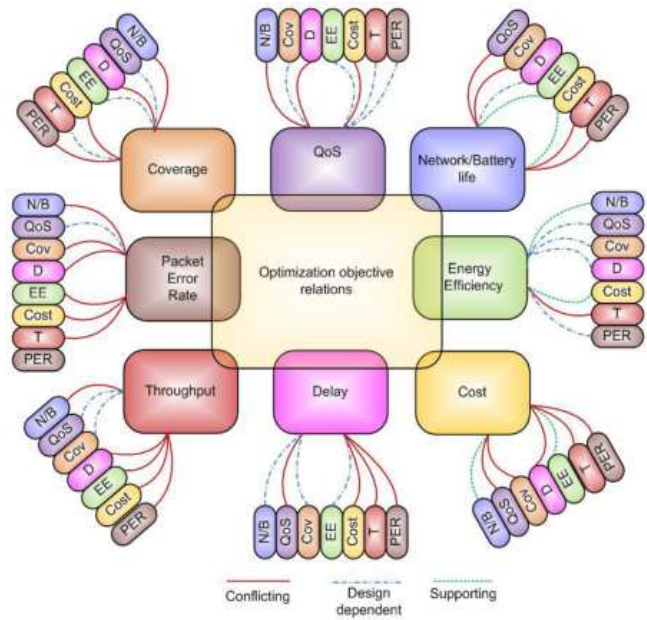

Figure 4. ECQSR objectives relationship

where: (N-B)=network-battery life; (QoS)=quality of service; (D)=delay; (EE)=energy efficiency; (Cost)=total cost of the system; (T)=throughput system; (Cov)=coverage; (PER)=packet error rate.

These different objectives may conflict or not with each other; Different objectives are associated together with various types of lines depending on the relationship between these objectives. The red line connects two objectives which are a conflicting relationship, for example, maximization the coverage conflicts with the delay, packet error rate, total cost of the system, and the network-battery lifetime. Whereas, the blue dashes and dots lines that connect two objectives mean that they have no direct relationship with each other [14].

The most important desired objectives of the ECQSR protocol are summarized as follows:
- To Keep the sensor nodes working as much as possible, Therefore, extending the lifetime of the network
- To Reduce the cost by selecting the shortest path which is produced by using nearest neighbor algorithm (NN), which meets the end to end delay requirements (minimizing the cost of the system)
- To Maximize the coverage and the throughput of the network.
- To Minimize energy consumption and packet error rate.

## 5. METHODOLOGY

This paper aims to make the routing between the sensor nodes more energy efficient, least cost and increasing QoS. Consequently, the network lifetime will be extended. To achieve these goals, the methodology of this paper based on the way of sensor nodes distribution, also using appropriate algorithms that help to make the technique of exchanging data between these nodes more flexible and shorter, these algorithms are the nearest neighbor algorithm (NN), and the simulated annealing (SA) algorithm.

These sensor nodes were distributed in the flat sensing filed according to certain coordinates (x, y), this data set which contains the sensor node coordinates was inserted into both algorithms (NN and SA) to find the shortest path of exchanging data between these sensors, each one of these algorithms has its special way to find this route. After implementing the two algorithms, simulation results show clearly that the NN algorithm attained better results than the SA algorithm. Section (7.3.3) shows the comparison results between the two algorithms.

## 6. PROPOSED ALGORITHM

Nearest Neighbour (NN) algorithm is used to find the optimal path between the sensors that were distributed in the sensing field taking into consideration of power consumption and cost reduction, the next section of this paper (Experimental Evaluation) will show the implementation of the algorithm with the simulation results.

## 7. EXPERIMENTAL EVALUATION

In this section, the experimental evaluation of ECQSR protocol will be shown, whether through the nearest neighbour, or simulated annealing algorithm, which will be implemented in MatLab version 7.9.0.529 R2009b 32-bit (win32).

### 7.1. Implementation

The simulation process of ECQSR protocol was evaluated and tested via nearest neighbor algorithm, which has been implemented using MatLab, to ensure the performance and the effectiveness of ECQSR protocol. Also, the simulated annealing algorithm was implemented in MatLab to make a comparison between the results of the two algorithms.

The Simulation process consists of 2000 nodes that are randomly distributed in some sensing field; all sensors are identical in battery source, memory, and a processing unit. Also, all these nodes were located in the same external environment.

The simulation results of the two algorithms have shown that the implementation of the ECQSR protocol with nearest neighbor algorithm has achieved a shorter path to travel between all nodes than simulated annealing, which means that the nearest neighbor will reduce the power consumption, cost reduction, and some other parameters. The simulation results of ECQSR protocol, whether using the nearest neighbor algorithm (NN) or simulated annealing algorithm, will be discussed in the next sections of this chapter.

**7.2. Data set**

To make an implement for the two algorithms we have to find a data set that represents the locations of the distributed node sensors, also must be compatible with the concept of the algorithms. The data set which is used in the implementation is organized as coordinate of a point, which are written as ordered pair as it is shown below, letter P represents the location of the node. The location of these nodes is determined using an integrated function that exists in MatLab.

P (14991 8390)

The simulation was applied at different locations of 2000 node sensors; all these sensors were distributed randomly on some sensing field.

**7.3. Simulation results**

After the implementation of the two algorithms, we obtained the following results; each result of these algorithms will be discussed separately:

**7.3.1. Implementation of nearest neighbor (NN) algorithm**

NN algorithm was implemented as a solution for finding the shortest path between all distributed node sensors; it has been implemented in two different programming languages:
a. MatLab:

Which achieved the desired results, with regard to finding the shortest path between all nodes, Figure 5 shows the execution result of NN on MatLab.

Figure 5. NN implementation on MatLab for length 730231.4981

The above figure shows the path between the nodes (2000 nodes) that are distributed in the sensing field using the nearest neighbor algorithm which is implemented in MatLab.

b. Java

NN also implemented by using Java that is also achieved a significant result, as shown in Figure 6.

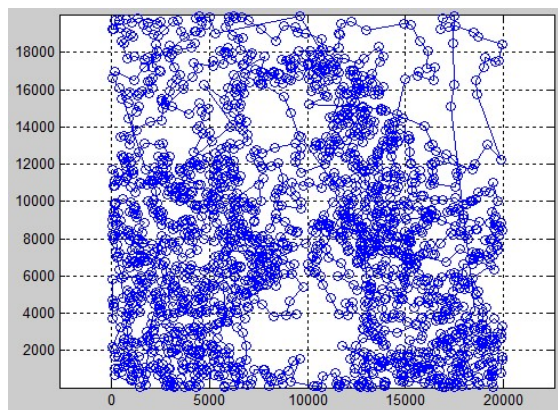

Figure 6: NN implementation in JAVA

The above figure illustrates the path between the nodes (2000 nodes) that are distributed in the sensing field using the nearest neighbor algorithm which is implemented in JAVA. Table 1 shows the implementation results of the NN algorithm in both PL (MatLab and Java).

Table 1. NN Implementation Result in MatLab and Java

| Algorithm | Programming language | COST | TIME |
|---|---|---|---|
| NN | MatLab | 730231.4981 | 3.66 Min |
|  | Java | 734095 | 25.60 Sec |

### 7.3.2. Implementation of simulated annealing (SA) algorithm

SA algorithm was implemented in MatLab; Figure 7 shows the execution result of SA on MatLab.

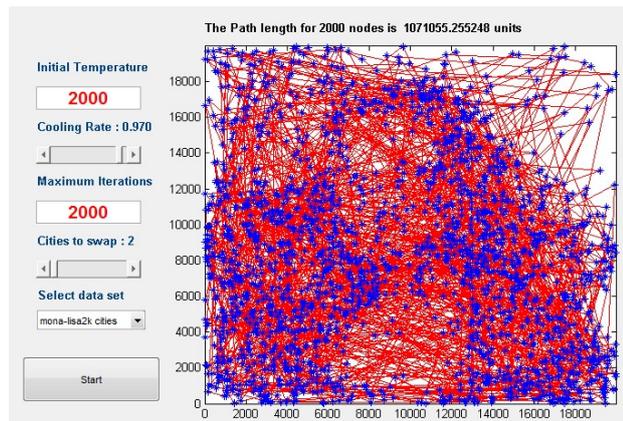

Figure 7. SA implementation on MatLab

The above figure shows the path between the nodes (2000 nodes) distributed in the sensing field using the simulated annealing algorithm which is implemented by MatLab; this figure should be compared with Figure 5. Table 2 shows the implementation results of the SA algorithm (MatLab).

Table 2. SA Implementation Result in MatLab

| Algo | Programming language | COST | TIME |
|---|---|---|---|
| SA | MatLab | 1071055.255 | 40 Sec |

When comparing Table 1 with Table 2, SA scored faster results than NN and higher cost.

### 7.3.3 Comparison between NN and SA algorithm according to path length

Figure 8 shows the average path-length of the ECQSR protocol, using the NN algorithm, and SA algorithm, it is clear to conclude that the SN is having a longer path than NN when the ECQSR is implemented.

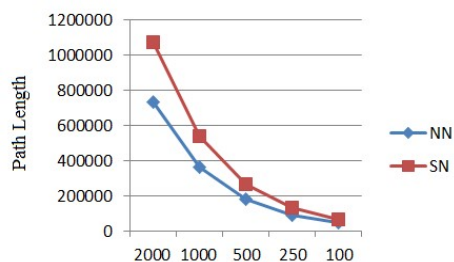

Figure 8: ECQSR Path-Length average

## 8. CONCLUSION

In this paper; Efficient Energy, Least Cost, quality of service based routing protocol for sensor networks protocol (ECQSR) has been presented specially for wireless sensor networks to extend the life of the network, finding lower cost by determining the shortest path using nearest neighbor algorithm (NN), put certain constraints on the delay of the path for real-time data, minimize the power consumption and error rate and other parameter.

Through the computer simulation, ECQSR multiple routing protocols have been implemented and evaluated, and it has proven its ability to maximize the performance of the WSN and extend the life of the network, through its capability in finding the shortest path which helps in reducing the power consumption, and cost reduction. As future research, it is intended to study the efficiency and the performance of ECQSR protocol, and analyze the effect of the path length, buffer size, and network size, on the metrics of the performance of ECQSR protocol.

## ACKNOWLEDGEMENTS

This research is funded by the Deanship of Research and Graduate Studies in Zarqa University, Jordan.

## REFERENCES

[1] Jalel ben-othman,bashir yahya, Energy efficient and QoS based routing protocol for wireless sensor networks, J. Parallel Distrib. Comput. 70, 849-857, 2010.
[2] X. Huang, Y. Fang, Multiconstrained QoS multipath routing in wireless sensor networks, Journal of Wireless Networks 14;(4), 465-478, 2008.
[3] A. Mainwaring, J. Polastre, R. Szewczyk, D. Culler, J. Anderson, Wireless sensor networks for habitat monitoring, in The Proceedings of the 1st ACM International Workshop on Wireless Sensor Networks and Applications, ACMWSNA, Atlanta, Georgia, USA, September 2828, pp. 8897, 2002.
[4] Anahit Martirosyan, Azzedine Boukerche, Richard Werner, Nelem Pazzi, Energy-aware and quality of service-based routing in wireless sensor networks and vehicular ad hoc networks, Annales des Telecommunications 63 (11-12) 669681, 2008.
[5] Suprotim Sinha Majumdar, Marut Pattanaik and JV Alamelu.Energy Efficient Wireless Sensor Network For Polyhouse Monitoring. European Journal of Advances in Engineering and Technology, 2(6): 77-82, 2015.
[6] Ghassan Samara, Mohammad Aljaidi, Aware-routing protocol using best first search algorithm in wireless sensor. The International Arab Journal of Information Technology, 15(3A): 592-598, 2018.
[7] Nida Maheen, Sanath Kumar, Efficient, Least Cost, Energy-Aware (ELCEA) Quality of Service Protocol in Wireless Sensor Networks, International Journal of Science and Research (IJSR), 2319-7064, 2014.
[8] Muhammad Iqbal, Muhammad Naeem, Alagan Anpalagan, Ashfaq Ahmed, Muhammad Azam, Wireless Sensor Network Optimization: Multi-Objective Paradigm, Sensors 2015, 20 July 2015, 15, 17572-17620; doi:10.3390/s150717572.
[9] Antonio Moschitta and Igor Neri. (2014), Power consumption Assessment in Wireless Sensor Networks.
[10] José-F. Martinez, Ana-B. Garcia, Ivan Corredor, Lourdes Lopez, Vicente Hernandez, Antonio Dasilva, Trade-Off Between Performance and Energy Consumption Wireless Sensor Networks, Lecture Notes in Computer Science (LNCS),4725/2007, Springer-Verlag, Berlin, Heidelberg, pp. 264271, ISBN:978-3-540-74916-5, 2007.
[11] H Makableh Ala'a, Ghassan Samara, Impact of Node Clustering on Power Consumption in WSN, The 7th International Conference on Information Technology, ICIT, 266, 269, 2015.
[12] Ye Ming Lu, Vincent W.S. Wong, An energy efficient multipath routing protocol for wireless sensor networks, International Journal of Communication System 20;(7), 747-766, 2007.
[13] Ghassan Samara, Khiri M Blaou, Wireless sensor networks hierarchical protocols,2017, Information Technology (ICIT), 2017 8th International Conference on, 998-1001.
[14] Liu,W.; Qin,G.; Li, S.; He, J.; Zhang, X. A Multiobjective Evolutionary Algorithm for Energy-Efficient Cooperative Spectrum Sensing in Cognitive Radio Sensor Network. International Journal of Distributed Sensor Networks, volume 2015, Article ID 581589, 2015.

## BIOGRAPHIES OF AUTHORS

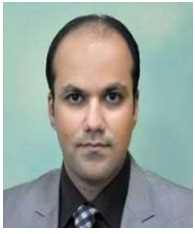

Dr. Ghassan Samara holds BSc. and MSc. in Computer Science, and PhD in Computer Networks. He obtained his Ph.D, from Universiti Sains Malaysia (USM) in 2012. His field of specialization is Cryptography, Authentication, Computer Networks, Computer Data and Network Security, Wireless Networks, Vehicular Networks, Inter-vehicle Networks, Car to Car Communication, Certificates, Certificate Revocation, QoS, Emergency Safety Systems. Currently, Dr. Samara is an assistant professor at the Computer Science Department, Zarqa University, Jordan.

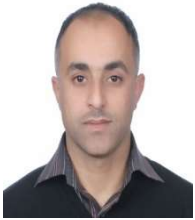

Mohammad Aljaidi received his BSc in Computer Science with honors degree from Zarqa University in 2014, and MSc in Computer Science with honors degree from Zarqa University in 2017. He is currently working as a system administrator in the Royal Jordanian Air Force (RJAF). His research interests include Wireless Sensor Networks (WSNs), Vehicular Adhoc Networks (VANETs), and Network Security.